\documentclass{aa} \usepackage{times} 
\usepackage{graphics} 
\usepackage{natbib} \bibpunct{(}{)}{;}{a}{}{,}

\def\lesssim{\mathrel{\hbox{\rlap{\hbox{\lower4pt\hbox{$\sim$}}}\hbox{$<$}}}}

\begin{document}

\title{Investigating the nature of the $z\simeq2.8$ submillimeter
selected galaxy SMM~J02399$-$0136 with VLT spectropolarimetry\thanks{
Based on observations collected at the European Southern Observatory,
Paranal, Chile (ESO Programme 64.P$-$0072) }}

\author{Jo\"el Vernet \inst{1,2} \and Andrea Cimatti \inst{2}}

\institute{European Southern Observatory, Karl Schwarzschild Str. 2,
D-85748, Garching bei M\"unchen, Germany \and Osservatorio Astrofisico
di Arcetri, Largo E. Fermi, I-50125, Firenze, Italy}

\offprints{Jo\"el Vernet, \email{vernet@arcetri.astro.it}}

\date{Received 8 August 2001 / Accepted 18 September 2001}

\abstract{We present deep optical spectropolarimety of
SMM~J02399$-$0136 ($z=2.8$) done with the VLT \emph{Antu} 8.2 m
telescope equipped with FORS1.  Moderate continuum and emission line
polarization are measured ($P\sim$5\%). We do not detect broad lines
in scattered flux as would be expected for a type-2 object but rather
a polarization behaviour similar to BAL quasars.  This classification
is confirmed by the detection of both high and low ionization broad
absorption troughs and a very red continuum. We argue that this object
shares several properties with local ULIGs such as Mrk 231 and other
ultraluminous infrared Lo-BAL quasars.  However, the fact that the
ultraviolet spectrum is dominated by non-stellar radiation does not
prove that the dust that is thermally radiating in the far infrared is
predominantly heated by the AGN. Since the energy that we get in the
far-infrared is precisely that which is removed from the ultraviolet
spectrum, this could mean that the starburst is more dust-enshrouded 
than the AGN due to a peculiar dust distribution. The
limits we place on the putative starburst contribution to the
restframe ultraviolet continuum together with constraints on the
amount of extinction provides an upper limit to the star formation
rate of about 2000 M$_{\odot}\,$yr$^{-1}$, consistent with previously
claimed high star formation rates level in this object.
}

\titlerunning{Investigating the nature of SMM~J02399-0136}
\authorrunning{J. Vernet \& A. Cimatti}

\maketitle

\section{Introduction}
Recent submillimeter observations revealed a population of dusty
ultraluminous infrared galaxies (ULIGs; $L>10^{12}L_{\odot}$) at high
redshifts (e.g. Smail et al. 1997\nocite{smail97}; Barger et
al. 1998\nocite{barger98}; Hughes et al. 1998\nocite{hughes98};
Cimatti et al. 1998\nocite{cimatti98c}). Understanding the nature of
such galaxies is important in order to verify if they are the
progenitors of the present-day massive spheroidals and whether massive
galaxies formed through an episode of rapid and strong star
formation. During such a starburst phase, a large amount of dust would
be produced and the rest-frame ultraviolet (UV) radiation of OB stars
would be absorbed by the dust grains and re-emitted in rest-frame
far-infrared (FIR), thus making these galaxies strongly extincted at
UV-optical wavelengths, but very luminous in the rest-frame FIR.

The inferred rest-frame FIR luminosities of the submm galaxies are
very high ($L>10^{12-13}L_{\odot}$), thus implying star formation
rates (SFRs) of the order of 500-2000 M$_{\odot}$yr$^{-1}$. The
presence of large reservoirs of molecular gas available to form stars
has been confirmed through the detection of CO emission in a few
submillimeter selected galaxies (M$_{H_2} \sim \times 10^{11}
$M$_{\odot}$; Frayer et al. 1998, 1999\nocite{frayer98b,frayer99};
Andreani et al. 2000\nocite{andreani2000}).

However, such high SFRs are usually derived assuming that the dust is
heated by young massive stars only and without any contribution from
an active galactic nucleus (AGN) possibly present. In fact, if a
quasar nucleus is hidden in the dust, its UV radiation could
significantly contribute to the FIR luminosities, and the SFRs would
be overestimated by large factors, thus leading to an incorrect
estimate of the contribution of these galaxies to the global star
formation history of the Universe (see Barger et
al. 2000\nocite{barger2000}). We recall here that the same problem is
present in low redshift ULIGs (see Sanders \& Mirabel
1996\nocite{sanders96} for a review).

In a first attempt to investigate the nature of high-$z$ submillimeter
selected galaxies and the link between AGN and starburst activity in
these sources, we observed SMM~J02399-0136, a $z=2.8$ galaxy
discovered as the counterpart of a SCUBA source found in a survey at
850$\mu$m of lensing cluster of galaxies (Ivison et
al. 1998\nocite{ivison98}[I98 hereafter]).  Its inferred unlensed FIR
luminosity is about $10^{13}$ L$_{\odot}$, placing this object in the
class of ``hyperluminous'' infrared galaxies, and the SFR estimated by
I98 was in the range of 2000-6000 M$_{\odot}$yr$^{-1}$.  Thanks to the
gravitational lensing of the cluster (whose amplification factor is
known to be 2.5), the flux from this galaxy is high enough to provide
a natural laboratory for detailed studies. For this reason, as a
``pilot'' study, we performed optical spectropolarimetry with the ESO
Very Large Telescope (VLT) with the main aim of studying the AGN and
starburst activities of this galaxy.

Throughout this paper we assume $H_0=50$ km~s$^{-1}$ Mpc$^{-1}$,
$\Omega_0=1$ and $\Omega_{\Lambda}=0$ unless otherwise stated.

\section{Observations and data reduction}

Spectropolarimetric observations were carried out using the imaging
spectrograph FORS1 (Appenzeller et al., 1992\nocite{appenzeller92}) in
PMOS mode at the VLT 8.2m unit telescope 1 (Antu) on December 1, 1999.
The detector is a Tek. 2048$ \times $2048 CCD with 24$ \mu m $ pixels
which correspond to a scale of 0.2\arcsec~pixel$ ^{-1} $.  The
polarization optics of the instrument are composed of a Wollaston
prism for beam separation and a rotating achromatic half-wave plate
mosaic.  A 1\arcsec~wide slit was used for all science
observations. We used the 300 lines mm$ ^{-1} $ grism (GRISM\_300V)
providing a dispersion of $ \sim $2.6~\AA~ pixel$ ^{-1} $ and an
effective resolution of about 11~\AA~FWHM. Observations were divided
into three sets of four exposures each of 1500 seconds with the
half-wave plate position angles set to 0\degr, 22.5\degr, 45\degr~ and
67.5\degr~ to reach a total exposure time of 5 hours. The slit was
oriented at 88.6\degr~ in order to include the two main components L1
and L2 of SMM~J02399$-$0136 (see I98). All observations were done
under subarcsecond seeing conditions with a seeing better than
0.6\arcsec~ during the last polarimetric set.

After debiasing and flat-fielding, the spectra were cleaned for cosmic
ray hits. The brightest component L1 is dominated by an unresolved
source and was extracted using a 1.8\arcsec~ aperture, whereas L2 is
more extended and was extracted using a 2\arcsec~ aperture. Wavelength
calibration was done using HgCd, He and Ar arc spectra. After accurate
wavelength registration using sky lines, the spectra were combined to
form the Stokes parameters Q and U following the method described in
Cohen et al. (1995)\nocite{Cohen95}.  Unbiased values for the
fractional polarization were estimated with the best estimator given
by Simmons \& Stewart (1985)\nocite{simmons85}. Statistical confidence
intervals on the fractional polarization and the polarization angle were
determined using a Monte-Carlo simulation taking into account the
effect background polarization and the detector noise (see Vernet
2001\nocite{vernet2000} for more details). We checked the polarimeter
observing the null polarization standard HD~64299. The measured
polarization in the B band is $ P=0.1295\pm 0.0063\% $, consistent
with $ P=0.151\pm 0.032\% $ given by Turnshek et
al. (1990)\nocite{turnshek90}. The polarization angle ($\theta , $
position angle of the E-vector) was corrected for effects of the
half-wave plate fast axis rotation with wavelength using calibration
data obtained during the instrument commissioning provided by ESO. The
position angle offset between the half-wave plate coordinate and the
sky coordinates was checked against values obtained for the polarized
standard HD~251204. Discrepant values for the polarization angle of
this star are reported in the literature.  We measure $\theta
=153.67\pm 0.055\degr$ in the V band, within $2\degr$ from the value
measured by Ogle et al. (1999)\nocite{ogle99} using the Keck
polarimeter (LRISp).  The flux calibration was done using observations
of the spectrophotometric standard star LTT 3218 (Hamuy et
al., 1992)\nocite{hamuy92}, and the atmospheric extinction correction
was done using CTIO extinction data since no extinction measurements
for the Paranal observatory are yet available. Finally, the total flux
spectra were corrected for Galactic extinction using $ A_{B}=0.135 $
mag. from Schlegel et al. (1998)\nocite{schlegel98} maps and the
extinction curve from Cardelli, Clayton \& Mathis
(1988)\nocite{cardelli88}.  Integrating the spectrum over the R band
gives R$ \simeq $22.5, close to the published value of 22.60$ \pm 0.04
$ given in Ivison et al. (1998)\nocite{ivison98}.

\section{Results}

\subsection{Polarization results}

\begin{figure}
{\par\centering
\resizebox*{1\columnwidth}{!}{\includegraphics{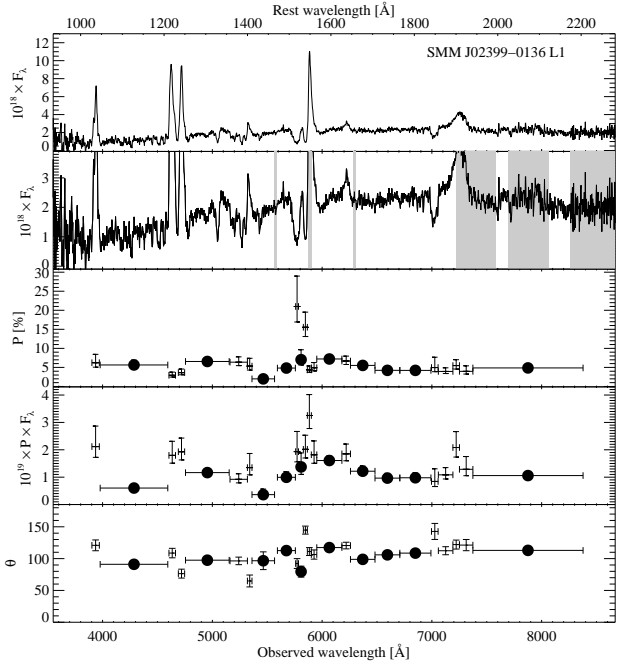}}
\par}

\caption{\label{pol_L1}Spectral and polarization properties of the
main component L1.  In each panel, \emph{from top to bottom:} the
observed total flux spectrum $ F_{\lambda }$ in units of
$10^{-18}\times $erg$\, $s$^{-1}\,$cm$^{-2}\,$\AA$ ^{-1}$ plotted on
two different scales, the first to show strong emission lines and the
second to show details of the continuum and absorption troughs, the
percentage polarization with 1$\sigma$ error bars, the polarized flux
in units of $10^{-19}\times $erg$\, $s$^{-1}\,$cm$^{-2}\,$\AA$^{-1}$
and the position angle of the electric vector with 1$ \sigma $ error
bars. Filled circles and crosses respectively indicate continuum bins
and emission lines bins with their underlying continuum. Shaded areas
indicate regions of strong sky emission. The spectrum has been
corrected for A and B band telluric absorptions.}
\end{figure}

Spectropolarimetry results for the brightest component L1 are
displayed in Fig.  \ref{pol_L1}. Data were binned to follow the main
spectral features and to avoid strong sky lines while keeping a
reasonable signal to noise ratio.  Moderate continuum polarization is
detected in all continuum bins. Continuum polarization measurement
with 1$ \sigma $ confidence interval in a single large bin in the line
free region between He\textsc{ii}$\lambda$1640 and
C\textsc{iii}]$\lambda$1909 ($6320\leq\lambda_{obs}\leq6990$~\AA)
gives $P=4.0\pm0.4\%$ and $\theta =102.2\pm 3.3$. Similarly between
N\textsc{v}$\lambda$1240 and Si\textsc{iv}$\lambda\lambda$1393,1402
($4760\leq\lambda_{obs}\leq5150$) we find $P=6.5\pm0.7\%$ and
$\theta=97.5\pm3\degr$. While the continuum polarization increases
slightly to the blue, the polarization angle is weakly dependent on
wavelength. Note however that the two extreme wavelength bins in
Fig. \ref{pol_L1} seem to indicate a rotation of $\theta $ of about
10\degr.

Ly$ \alpha $ and N\textsc{v} lines show a significantly lower but
still significant polarization than the neighboring continuum. After
correction for underlying continuum polarization we find $P_{Ly\alpha
}=2.1^{+0.9}_{-0.5}\%$ and $P_{NV}=2.6_{-0.6}^{+1.0}\%$. Polarization
does not vary significantly within measurement errors across the
Si\textsc{iv}, C\textsc{iv}$\lambda\lambda$1548,1550, He\textsc{ii}
and C\textsc{iii}{]} lines (note however that C\textsc{iv} line
polarization is uncertain because it is affected by the relatively
strong 5890~\AA~ Na D atmospheric emission line).

We have measured the polarization in the two broad troughs blueward of
C\textsc{iv}.  We find $P=17^{+4}_{-3}\% $ and $ P=13^{+4}_{-2}\% $ in
the bluest and in the reddest trough respectively, significantly
higher than the neighboring continuum. As a result, the absorption
features are absent in the polarized spectrum (see Fig. \ref{pol_L1}). We
measure a C\textsc{iv} line FWHM in polarized flux of $\sim1930\pm730$
km~s$^{-1}$, within the range of line width measurements for other
strong lines in total flux.  However, given the low signal to noise
level within the small bins used to measure $P$ in the troughs, such a
result should be regarded with caution.

A consequence of the moderate polarization is the rather low signal to
noise on the polarization measurements for such a faint target which
forces us to apply a coarse binning to the data. This renders difficult
the detection of shallow features such as broad emission lines in
polarized flux. We find no evidence for the presence of broad emission
line features in polarized flux in our data.

\begin{figure}
{\par\centering
\resizebox*{1\columnwidth}{!}{\rotatebox{90}{\includegraphics{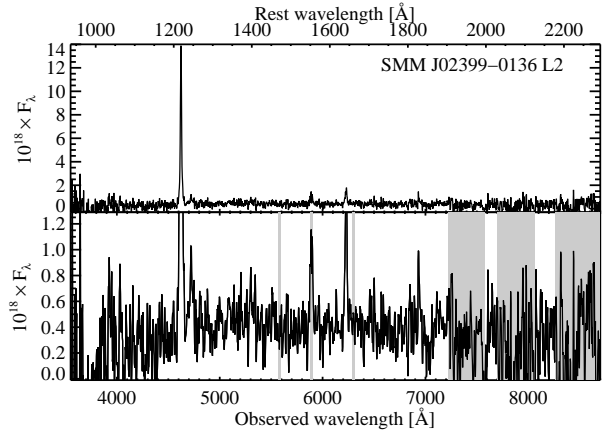}}}
\par}

\caption{\label{pol_L2}Spectrum of the L2 component in units of $
10^{-18}\times $erg$\,$s$^{-1}\,$cm$^{-2}\, $\AA$ ^{-1} $ plotted on
two different scales. The spectrum in the bottom panel has been
smoothed using a 3 pixels boxcar.}
\end{figure}
Polarization measurements for the component L2 are much more difficult
and uncertain since it is about two magnitudes fainter than L1. We
estimated the continuum polarization in a single large bin giving
$P=9\pm3\%$ (3$ \sigma $ interval) and $ \theta =140\degr \pm 4\degr $
(1$ \sigma $ error). This values should however be considered with
caution because at such a low signal to noise level any residual of
cosmic hits or imperfect bright sky line subtraction can have dramatic
effects on polarization measurements.
%\vspace{0.3cm}

\subsection{Total flux spectrum}

A consequence of the long integration time required by
spectropolarimetric observations is the obtaining of a very deep total
flux spectrum (S/N$ \sim $15 per continuum resolution element for L1)
allowing for the detection of several previously undetected absorption
features in L1 and emission lines in L2.  The total flux spectrum of
L1 is presented on the top two panels in Fig. \ref{pol_L1} and the
spectrum of L2 in Fig. \ref{pol_L2} on two different scales.

Component L1 shows a very red continuum with a slope $ \beta _{1500}=
0.6\pm 0.2 $ (with $ F_{\lambda }\propto \lambda ^{\beta } $ measured
between 1300$<\lambda_{rest}<$1800~\AA). The shape of the continuum is
rather complex especially between Ly$ \alpha $ and C\textsc{iv}
probably due to blending of several absorption and emission
features. The continuum of L2 is also well detected and is roughly
flat in $ F_{\lambda } $ ($ \beta \sim 0 $).

\begin{figure}
{\par\centering
\resizebox*{1\columnwidth}{!}{\rotatebox{0}{\includegraphics{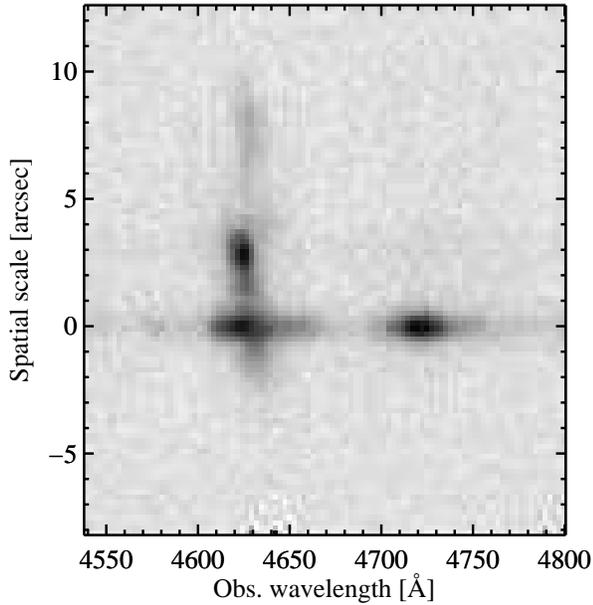}}}
\par}

\caption{\label{figlya}The 2D-spectrum of SMM~J02399-0136 in the
region of the Ly$\alpha$ and N\textsc{v} emission lines. The origin of
the spatial scale is set at the continuum peak of L1. The Ly$\alpha$
emission of the L2 component is clearly visible about 3\arcsec~away
from L1,but the continuum is very weak. The total extent of the
Ly$\alpha$ halo is about 13\arcsec.}
\end{figure}

Emission line measurements and identifications for each component are
given in table \ref{emission}. As reported in I98 and
Villar-Mart\'{\i}n et al. (1999)\nocite{villar99}[VM99 herafter], the
Ly$\alpha$ line emission extends over 13\arcsec, far beyond the extent
of the detectable continuum (see Fig. \ref{figlya}).  We do not detect
extended emission in other strong emission lines. The observed
emission line properties of L1 are consistent with the ones found in
the spectra presented in I98 and VM99, except for the Ly$ \alpha $
which is significantly stronger in VM99 probably because their L1
spectrum was extracted using a wider aperture. The strong emission
lines have a typical width of $ \sim $2000 km~s$^{-1}$.  Si\textsc{iv}
and C\textsc{iv} show strongly asymmetric profiles due to absorption
troughs of their blue wing. Ly$ \alpha$, C\textsc{iv} and N\textsc{v}
line profiles do not show evidence of broad wings. C\textsc{iii{]}} is
remarkably broad compared to other lines, probably due to blending with
Fe\textsc{iii} UV 34 $\lambda\lambda1895,1914,1926$ and
Si\textsc{iii}]$\lambda\lambda 1882,1892$ (see discussion).

\begin{figure}
{\par\centering
\resizebox*{0.90\columnwidth}{!}{\rotatebox{0}{\includegraphics{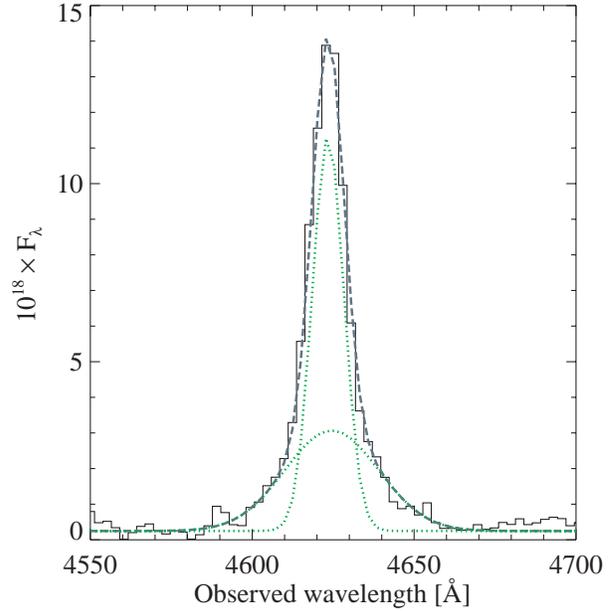}}}
\par}

\caption{\label{fig_L2lya}Two components fit to the Ly$\alpha$
emission line profile in L2.  Individual components are shown with
dotted lines. The dashed line represents the total fit and the
observed spectrum is displayed as a histogram. Parameters of the fit
are given in table \ref{emission}.  }
\end{figure}

Strong emission lines are also well detected in L2. The line ratios
are very different from the ones observed in L1, in particular
Ly$\alpha$/C\textsc{iv} and He\textsc{ii}/C\textsc{iv} are much larger
(see table \ref{emission}).  The typical line width is of the order of
1000 km~s$^{-1}$, narrower than in L1. The Ly$\alpha$ line shows a
broad (2200 km~s$^{-1}$) and a narrow (750 km~s$^{-1}$) component. We
show in Fig. \ref{fig_L2lya} a two component fit of this line.  Note
that the width of the broad component is similar to line widths
measured in L1 (see table. \ref{emission}).

We computed the redshift based on measurements of the
He\textsc{ii}$\lambda$1640 line rather than other stronger emission
lines because this recombination line is not affected by absorption.
We find $z=2.7947\pm0.0004$ for L1, consistent with the redshift
measured from the [O \textsc{iii}]$\lambda\lambda$4959,5007 in near
infrared spectra by I98.  Measurements for L2 give $z=2.7981\pm0.0003$
placing this second component about $270$km~s$^{-1}$ from L1, also
consistent with results found by I98 based on rest-frame optical
lines.

\begin{figure}
{\par\centering
\resizebox*{0.98\columnwidth}{!}{\includegraphics{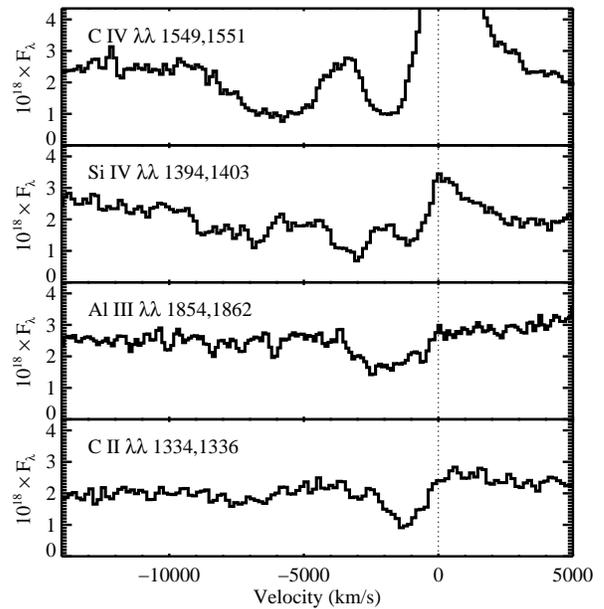}} \par}

\caption{\label{velprof}Velocity profiles of the
C\textsc{iv}$\lambda\lambda$1549,1550,
Si\textsc{iv}$\lambda\lambda$1394,1403,
Al\textsc{iii}$\lambda\lambda$1854,1862,
C\textsc{ii}$\lambda\lambda$1334,1336 absorption troughs. The zero
velocity is taken at the expected emission wavelength of the reddest
component of the doublet, computed using the redshift determined with
the He\textsc{ii} line.}
\end{figure}

One striking property of the spectrum of L1 is the presence of broad
absorption lines (BAL). The presence of high ionization Si\textsc{iv}
and C\textsc{iv} BAL was already noted by I98. We report here the new
detection of low ionization absorption lines of Al\textsc{iii}$ \lambda
\lambda $1854,1862, C\textsc{ii}$ \lambda \lambda $1334,1336,
Si\textsc{ii}$ \lambda\lambda $1304,1309 and Si\textsc{ii}$ \lambda
\lambda $1260,1265.  Observed wavelength and equivalent widths of the
low ionization absorption lines are given in table \ref{low_ion_abs}.
Velocity profiles of the strongest absorption troughs are shown in
Fig.  \ref{velprof}. While high ionization lines have two main
velocity complexes at $-$1000 and $-$6700 km~s$^{-1}$, low ionization
absorption lines are only detected in the lowest velocity
subcomponent.

\begin{table*}
{\centering \begin{tabular}{lllccccl} Comp.& Id.& $\lambda_{vac}$ & $
\lambda _{obs} $ & $ F_{\lambda }\times 10^{17} $ & $
W_{\lambda}^{obs} $ & FWHM & Comments\\ & & (\AA)&(\AA)& ($ $erg$\,
$s$^{-1}\, $cm$^{-2}$)& (\AA)& ($ $km$\: $s$^{-1} $)& \\ \hline \hline
L1& O \textsc{vi} &1031.9,1037.6& 3938.2$ \pm $0.2& 20.8$ \pm $1.2&
240$ \pm $53& 2367& \\ & Ly$\alpha$ &1215.7 & 4628.9$ \pm $0.1& 34.4$
\pm $0.3& 235$ \pm $7& 2383& 2 components\\ & N \textsc{v}
&1238.8,1242.8& 4720.0$ \pm 0.1 $& 28.2$ \pm $0.4& 205$ \pm $9& 1861&
2 components\\ & Si \textsc{iv} &1393.8,1402.8& 5334.6$\pm$0.1&
3.7$\pm$0.2& 21$\pm$1& 1899& blue wing absorbed\\ & C \textsc{iv}
&1548.2,1550.8& 5891.8$\pm$0.1& 27.4$\pm$0.6& 124$\pm$7& 1491& blue
wing absorbed \\ & He \textsc{ii} &1640.4 & 6223.3$ \pm$0.6 & 4.3$ \pm
$0.4& 17.2$ \pm 1.7 $& 2330& \\ & C \textsc{iii}{]} &1906.7,1908.7&
7247.9$\pm$0.5& 27.5$\pm$0.7& 122$\pm$8& 4672& probably blended with\\
%&&&&&&Fe\textsc{iii} UV 34 $\lambda\lambda1895,1914,1926$\\
%&&&&&&&and Si\textsc{iii}]$\lambda\lambda 1882,1892$\\
&&&&&&&Fe\textsc{iii} UV 34 and Si\textsc{iii}]\\ &&&&&&&lines (see
text)\\ \hline L2& Ly$\alpha$ narrow&1215.7& 4623.4$ \pm $0.02& 13.6$
\pm $1.3& 546$ \pm $22& 744.2&\\ & Ly$\alpha$ broad&1215.7& 4624.6$
\pm $0.7& 10.2$ \pm $1.4& 419$ \pm 103 $& 2212&\\ & N \textsc{v}
&1238.8,1242.8& 4723.3$ \pm $0.8& 1.64$ \pm $0.5& 40$ \pm 17 $& 2026&
broad and narrow comp.?\\ & C \textsc{iv} &1548.2,1550.8& 5891.5$ \pm
$0.1& 1.87$ \pm$0.1& 45.3$ \pm $5.6& 1385& strong sky emission\\ & He
\textsc{ii} &1640.4& 6228.9$ \pm $0.5& 2.48$ \pm $0.2& 59.0$ \pm $8.4&
963 & \\
%& Si\textsc{iii}{]} &1882.5,1892.0 & 7127$ \pm $1.2&0.95$ \pm $0.1& 30$ \pm $3& 51$ \pm $3&\\
%& C \textsc{iii{]}} &1906.7,1908.7& 7240$ \pm $1.5& 1.6$ \pm $0.2& 53$\pm $8& 47$ \pm $5&
\hline
\end{tabular}\par}

\caption{\label{emission}Emission line measurements.}
\end{table*}

\begin{table}
{\centering \begin{tabular}{llcc} \hline Identification& \( \lambda
_{vac} \) (\AA)& \( \lambda _{obs} \) (\AA) & \( W^{obs}_{\lambda } \)
(\AA) \\ \hline \hline ~Si \textsc{ii} &~1260.4,1264.7 &4780 &9\\

\( \begin{array}{l} \rm{O~\textsc{i}}\\ \rm{Si~\textsc{ii}}
\end{array}  \)			&
\( \left. \begin{array}{l} 1302.2,1304.9\\ 1304.4,1309.3
\end{array}\right\}  \)		&
4948& 8\\

~C \textsc{ii} &~1334.5,1335.7 &5048 &17\\ ~Al \textsc{iii}
&~1854.7,1862.8 &7021 &23\\ \hline
\end{tabular}\par}

\caption{\label{low_ion_abs}Low ionization absorption line
measurements in L1.}
\end{table}

\section{Discussion}

\subsection{The origin of the polarization}

Since this object suffers from very little Galactic extinction ($
E_{B-V}=0.031, $ using the Schlegel et al. (1998)\nocite{schlegel98} map
and assuming $ R_{V}=3.1 $), we are confident that the observed
polarization is intrinsic to SMM~J02399$-$0136 (we find a limit to the
interstellar polarization $P\lesssim 0.28\,\%$ using the Serkowski
(1975)\nocite{serkowski75} formula $P\leq9\times E_{B-V}\%$).  Thus, the
two mechanisms that can produce the observed level of both continuum
and emission line polarization in the rest-frame ultraviolet are
transmission of the radiation through magnetically aligned dust grains
(dichroic polarization) within SMM~J02399$-$0136 and scattering by
electrons and/or by dust.

The possibility of dichroic polarization can be excluded considering
the shape of $ P(\lambda $). The variation of $ P(\lambda $) at
ultraviolet wavelengths for polarization by transmission is well
parameterized by an extended Serkowski law (Martin, Clayton and Wolff,
1999\nocite{martin99}) obtained by fitting observations of different
lines of sight in our Galaxy. It is has a characteristic shape with a
rather steep increase to the red and a maximum between 4000 and 7000~
\AA. Unless the polarization of SMM~J02399$-$0136 is diluted by an
extremely red unpolarized continuum component or its dust properties
are very different from the Galactic ones, the increase of $ P(\lambda
)$ to the blue in L1 is not compatible with polarization by
transmission. Therefore, we consider scattering by dust and/or
electrons as the most likely dominant polarization mechanism.

\subsection{The nature of the AGN}

The two classes of objects that commonly show a significant amount of
scattered light in their UV spectrum are type-2 AGN (i.e. Seyfert 2
and radio galaxies, see e.g. Antonucci \& Miller
1985\nocite{antonucci85}, di Serego Alighieri et
al. 1989\nocite{alighieri89}) and broad absorption line quasars (see
e.g. Goodrich \& Miller 1995\nocite{goodrich95a}). In type-2 objects,
scattering is produced by material distributed on the kpc scale
illuminated by the anisotropic radiation from an active nucleus
surrounded by an optically thick torus that completely blocks the
direct view to the quasar (see e.g. review by Antonucci 1993). The
continum of these objects is usually well resolved in the rest-frame
UV and they typically show high continuum polarization and broad lines
detected in scattered flux (up to 20\%, see e.g.  the spectropolarimetric
study of $z\sim2.5$ radio galaxies by Vernet et
al. 2001\nocite{vernet2001}). In contrast, scattering is believed to
occur much closer to the nucleus in BAL QSOs, while the unpolarized
direct quasar radiation is less attenuated, resulting in a much
smaller range in observed polarization (see e.g. spectropolarimetric
study of a sample of BALQSO by Ogle et al. 1999\nocite{ogle99}[O99
hereafter]).

The detection of moderate continuum polarization together with the non-detection 
of broad emission lines in scattered flux and the presence
of broad absorption troughs suggests that L1 is more similar to a BAL
QSO than to a type-2 object.  This interpretation is strengthened by
the fact that L1 is dominated by a point source (the continuum
emission is unresolved in our spectrum, even in data obtained with
seeing $\sim0.5\arcsec$; I98 measure a FWHM of 0.1-0.2\arcsec~for L1
on HST WF/PC F702W and WFPC2 F336W images).  The presence of low
ionization absorption further identifies L1 as one of the very few low
ionization BAL QSOs known (Lo-BAL QSO). While about 12\% of known
quasars show high ionization BAL (Weymann et al.
1991\nocite{weymann91}), a small subset of these (about 15\%,
Hutsemékers et al. 1998\nocite{hutsemekers98}) also show low
ionization broad absorption lines like Mg\textsc{ii$ \lambda \lambda
$2796,2804} and in some cases Al\textsc{iii}$ \lambda \lambda
$1854,1862.

The slow increase of the continuum polarization to the blue that we
detect in L1 is very similar to what is commonly observed in BAL QSO
(see O99).  The polarization level of L1 is clearly sitting in the
high tail of the polarization distribution of BAL QSO. This may
however be not too surprising considering that Lo-BAL QSOs seem to
show a wider range in polarization than high ionization BAL QSOs as
suggested by Hutsemékers et al. (1998)\nocite{hutsemekers98}.  Note
that the strong rise in polarization that we possibly detect in the
C\textsc{iv} absorption troughs is also a typical feature of BAL QSOs
(e.g. O99).  The very red color of the continuum ($ \beta
_{1500}\simeq 0.6 $) compared to normal quasars ($ \beta _{1500}\simeq
$-1.4, measured on Brotherton et al. 2001\nocite{brotherton2001}
composite quasar) is also a common feature observed in Lo-BAL
QSO. Measuring the UV continuum slope on Brotherton et
al. (2001)\nocite{brotherton2001} Lo-BAL QSO composite spectrum we
find $ \beta _{1500}\simeq 0.4 $, rather close to the color of L1.

The width of the lines of L1 ($\sim 2000 $ km~s$^{-1}$ FWHM) appears
to be intermediate between that of quasars (typically $ \sim
$5000 km~s$^{-1}$) and type-2 objects ($ \sim $1000 km~s$^{-1}$) while
the line ratios are more typical of a quasar. Emssion line modeling
performed by VM99 suggests that the emission line spectrum of L1 might
be dominated by an intermediate density region
($n_{H}\simeq10^{4}-10^{6}$).  The only line that appears to have a
broad and rather complex profile is C\textsc{iii}{]}$\lambda$1909.
Such a phenomenon has already been observed both in BAL QSO (Hartig \&
Baldwin 1986\nocite{hartig86}, Weymann et al. 1991\nocite{weymann91})
and narrow line quasars (e.g. Baldwin et al. 1988\nocite{baldwin88}).
In these objects, the apparent broad feature has usually been
interpreted as a blend of C\textsc{iii}{]}$\lambda 1909$,
Fe\textsc{iii} UV 34 $\lambda\lambda1895,1914,1926$ multiplet and
Si\textsc{iii}]$\lambda\lambda 1882,1892$ doublet.

Properties of L2 are significantly different from those of L1. Line
widths and line ratios are more typical of type-2 objects. This
together with the detection of a broader Ly$ \alpha $ component of a width
similar to that of L1 and possible high continuum polarization
suggests that the spectrum of L2 could be dominated by scattered light
from the AGN present in L1, similar to the extended UV emission
observed in high redshift radio galaxies.

SMM~J02399$-$0136 shows quite different properties from the three high
redshift hyperluminous infrared galaxies F~10214+4724, P~091004+4109
and F~15307+3252 that have been proven to be type-2 objects (Goodrich
et al. 1996\nocite{goodrich96}; Tran et al. 2000\nocite{tran2000};
Hines et al. 1995\nocite{hines95}). In fact, SMM~J02399$-$0136 shares
several of its unusual properties with the low redshift ULIG Mrk~231
that has also been classified as a Lo-BAL QSO (e.g. Smith et
al. 1995\nocite{smith95}). The nature of Lo-BAL QSO and in particular
their link with ULIGs is not clear at present.  It has been suggested
that some of these objects could be AGN/massive starburst composite
(Lipari et al., 1994\nocite{lipari94b}) or young quasars casting off
their cocoons of gas and dust (Sanders et al., 1988\nocite{sanders88},
Voit et al., 1993\nocite{voit93}).

\subsection{Constraints on the star formation activity}

The strong CO(3$ \rightarrow $2) emission and the inferred large mass
of molecular gas together with the high far-infrared luminosity
indicate that star formation is occurring at a rate greater than $
\sim 10^{3}$M$_{\sun }\,$yr$^{-1} $ (I98; Frayer et
al. 1998\nocite{frayer98}). Although dilution of the polarization by
starlight is not required by our spectropolarimetric data, a
significant contribution of a population of hot stars to the UV
continuum is not at all ruled out.

\subsubsection{Searching for starburst spectral signatures}
The two unambiguous direct spectral signatures of the presence of
young massive stars that we might expect to detect in the rest-frame
UV are Si\textsc{iv}$ \lambda $1400, C\textsc{iv}$ \lambda $1549
P-Cygni profiles and purely photospheric absorption lines (i.e. not
contaminated by interstellar absorption) from O and B stars such as
Si\textsc{iii$ \lambda$}1294, C\textsc{iii$ \lambda $}1427,
S\textsc{v}$ \lambda $1502 and N\textsc{iv}$ \lambda $1718.

Both the shape and the velocity extent of the troughs blueward of
C\textsc{iv} and Si\textsc{iv} emission lines are not compatible with
these features being dominated by stellar P-Cygni profiles.  While we
observe double absorption troughs that extend up to about $-$9000
km~s$^{-1}$ with rather sharp edges both in C\textsc{iv} and
Si\textsc{iv} lines (see Fig. \ref{velprof}) in L1, stellar wind
absorption troughs are in general strongly asymetric and extend no
further than about $-$5000 km~s$^{-1}$ in both modeled starburst
spectra (e.g. Starburst~99) and observed star-forming galaxies (see
e.g. local starburst galaxies templates NGC~1705, NGC~1741, NGC~4214
and IRAS~0833$+$6517 from Heckman \& Leitherer,
1997\nocite{heckman97}; Conti et al., 1996\nocite{conti96}; Leitherer
et al., 1996\nocite{leitherer96}; Gonzalez Delgado et al.,
1998\nocite{gonzalez98} respectively and also the $z=2.7$ lensed Lyman
break galaxy MS~1512$-$cB58 from Pettini et al. 2000).

We detect one absorption line at $\lambda_{obs}=5434.4 $~\AA~
($\lambda_{rest}=1432.1 $~\AA) relatively close to the expected
wavelength of the C\textsc{iii$ \lambda $}1427 photospheric
line. However, its equivalent width ($W^{rest}_{\lambda
}=1.75\pm0.1$~\AA~in the rest-frame) is significantly larger than the
maximum equivalent width of $\sim1$~\AA~ that we measured for this
line on both the above listed starburst templates and Starburst 99
models.  One more likely identification for this feature is that of a
Mg\textsc{ii}$\lambda\lambda$2796,2803 doublet due to an intervening
absorber.  A two gaussian fitting of this absorption feature yield
$\lambda_{obs}=5431.7$~\AA~ and $\lambda_{obs}=5445.8$~\AA~ with an
error of $\sim$0.5\AA, consistent with a
Mg\textsc{ii}$\lambda\lambda$2796,2803 doublet at $z=0.9424\pm0.0002$.

\subsubsection{Limits on the stellar continuum}
Since we did not find any direct spectral evidence for the presence of
a massive starburst in our spectrum, we used two different methods to
constrain the fraction of stellar continuum to the total flux
($f^{\star}$).

\paragraph{First method.}
We first computed an upper limit to the equivalent width of an
unresolved line at the expected wavelength of the four purely
photospheric lines listed above. We found the most stringent limit for
the N\textsc{iv}$\lambda$1718 line: in the region of the expected
observed wavelength ($\sim$~6521\AA), the continuum signal to noise
ratio is $\sim16$ which gives a 2$\sigma$ upper limit on the
rest-frame equivalent width of the N\textsc{iv}$\lambda$1718 line of
$\sim0.4$ \AA.  The equivalent width of this photospheric absorption
line measured on both continuous star formation and instantaneous
burst Starburst 99 model spectra (Salpeter IMF, solar metallicity)
ranges from 0.6 to 0.7 \AA~ for bursts ages $t^{\star}$ between 1 and
5 Myr.  This value is comparable to what is typically measured in
local starbursts (e.g. in Heckman \& Leitherer, 1997\nocite{heckman97}
HST/GHRS spectrum of NGC~1705 we measure
$W_{\lambda}$(N\textsc{iv})$\sim0.6$).  These numbers together with
our upper limit on the observed equivalent width allow us to put an
upper limit to the contribution at 1718~\AA~ of a young
($t^{\star}\leq5$ Myr) starburst of about 70\% of the total flux
($f^{\star}(1718)\lesssim70$\%).

\paragraph{Second method.}
Since we showed that the observed absorption troughs are intrinsic to
the active nucleus, we can also place an upper limit to $f^{\star}$ by
saying that the stellar continuum level has to be lower than the flux
at the center of the absorption troughs (that is to say that if these
BAL troughs were saturated then the stellar flux would be precisely
equal to the flux at the bottom of the troughs).  Considering the
deepest C\textsc{iv} trough we obtain that the stellar continuum flux
$ F_{\lambda}^{\star}(1500)$ at $ \sim $1500~\AA~ must be lower than
$9.5\,10^{-19}$erg$\, $s$^{-1}\, $cm$^{-2}\, $\AA$^{-1}$.  At
$\lambda_{rest}$=1500~\AA, the measured total flux is $\sim
2.3\,10^{-18}$erg$\, $s$^{-1}\, $cm$^{-2}\, $\AA$^{-1}$.  This then
gives a more srtingent limit than the previous method:
$f^{\star}(1500)\lesssim40$\%.

\paragraph{Polarization level consistency check.}
The lack of knowledge about the intrinsic AGN continuum polarization
level prevents us from putting meaningful constraints on $f^{\star}$
using polarization measurements. We can however reversely compute, as
a check, the limit on the intrinsic AGN polarization $P_{AGN}$
given $f^{\star}$: calling $P_{obs}$ the observed polarization,
$P_{AGN}$ is given by $P_{AGN}=P_{obs}/(1-f^{\star})$.  For
$f^{\star}(1500)\lesssim40$\%, taking $P_{obs}\sim5\%$ we find
$P_{AGN}\lesssim8\%$.  This value is higher than the avarage BAL QSO
continuum polarization but is comparable to what is observed in most
polarized objects in O99's sample (e.g. $P=5.41\%$ between 1600~\AA~
and 1840~\AA~ in the BAL QSO 1333$+$2840). Note also the this
continuum polarization level is still moderate compared to that of the
most polarized Lo-BAL quasar known FIRST~J15633.8$+$351758 for which
$P\simeq13\%$ near 2000~\AA~(Brotherton et
al. 1997\nocite{brotherton97}).

\subsubsection{Limits on the star formation rate}
We can now translate the most stringent limit we obtained on
$F_{\lambda}^{\star}(1500)$ in the observed frame into a limit on the
SFR.  The luminosity distance at $z=2.794$ for the cosmology we assume
in this paper is $D_L=6.83\,10^{28}$ cm.  The limit on the rest-frame
stellar luminosity at 1500~\AA~ is then
$L_{\nu}^{rest}(1500)\lesssim\,1.6\,10^{29}$erg$\,$s$^{-1}\,$Hz$^{-1}$.
Taking the Madau, Pozzetti \& Dickinson (1998)\nocite{madau98} conversion
factor between SFR and luminosity at 1500~\AA~
($L_{\nu}(1500)=8\,10^{27}\times \rm{SFR}/[M_{\sun}\,yr^{-1}]$) we
obtain an upper limit to the SFR of $\sim20\,$M$_{\sun}\,$yr$^{-1}$.
If we take into account the lensing factor of 2.5 given by I98 we have
SFR$\lesssim8\,$M$_{\sun}\,$yr$^{-1}$.

We know that star formation is strongly affected by dust extinction,
and we need to correct for this effect in order to set meaningful
limits on the SFR.  Any unpolarized stellar component that would
contribute significantly to the continuum ($f^{\star}\geq10\%$) would
affect the observed dependance of the polarization on wavelength if
its color is different from that of the AGN.  In L1, we know that
$P(\lambda)$ rises weakly to the blue. Since this behavior is very
similar to what is observed in most of the BAL QSO (see O99), we can
assume that the amount of dilution by starlight does not vary
significantly over the observed wavelength range which means that the
stellar component must have a spectral slope similar to that of the
total flux ($\beta _{1500}\sim 0.6$).  This slope is much redder than
the one predicted by stellar population synthesis models for young (a
few $ 10^{7} $yr) populations indicating that star formation is
heavily dust enshrouded, as suggested by the huge FIR luminosity of
this object.  Starburst~99 stellar population synthesis models predict
a slope $\beta _{1500} $ of about $-2.5$ for a $t^{\star}\geq10$ Myr
continuous burst (i.e. when the stellar population has reached an
equilibrium after the initial onset of the burst) which gives a
continuum slope index variation due to reddening $\Delta\beta
_{1500}\sim 3.1$. Assuming a Calzetti (1997)\nocite{calzetti97b}
extinction law we compute the slope index variation $\Delta\beta
=-5.9\times E_{B-V}$ and the extinction in magnitude at 1500~\AA~ $
A_{1500}=11.4\times E_{B-V} $ as a function of reddening.  This then
gives $ E_{B-V}\sim 0.52 $ and $ A_{1500}\sim 6.0 $ mag, an extinction
of a factor $ \sim $250 at 1500~\AA.

Combining this estimate with our limit on the SFR we obtain an
extinction corrected upper limit on the SFR of
$\sim2000$M$_{\sun}\,$yr$^{-1}$.  How sensitive is this limit to our
assumptions? Unless the observed absorption troughs do not originate
from within the AGN, the limit on the stellar flux at 1500~\AA~ is
quite reliable. This is in fact a rather conservative upper
limit. However, our estimate of the extinction, based on
considerations on polarization dilution by starlight, is valid only if
the stellar continuum makes up a significant fraction of the total UV
continuum. If for instance $f^{\star}$ is lower than $\sim10\%$ at
1500~\AA, the limit on the SFR before extinction correction becomes as
low as $2\,$M$_{\sun}\,$yr$^{-1}$. In this case we cannot constrain
the extinction with the present data.  It would then require an
extinction factor of $\sim1000$ ($A_{1500}\sim7.5$ mag.) to reach the
above limit of $2000\,$M$_{\sun}\,$y$r^{-1}$ or a factor $\sim3000$
($A_{1500}\sim8.7$ mag.)  to reach the total SFR of
$\sim6000\,$M$_{\sun}\,$yr$^{-1}$ derived by I98 from submillimeter
data.

\section{Conclusions}

We have presented in this paper deep low resolution optical
spectropolarimetry of the $ z\simeq2.8 $ submillimeter selected galaxy
SMM~J02399$-$0136. The main new observational results are the
following:

\begin{itemize}
\item The main component L1 shows moderate ($ \sim 5\% $) continuum
polarization. We also detect formally significant polarization of the
second component L2. We do not find any evidence for the presence of
broad scattered lines.
\item We do not detect any unambiguous signature of massive stars like
P-Cygni profiles or photospheric absorption lines in the rest-frame UV
spectrum of this object.
\item In addition to the already known high ionization absorption
troughs, we discovered broad low ionization absorption lines
(Al\textsc{iii}, C\textsc{ii} and Si\textsc{ii}).
\end{itemize}

While SMM~J02399$-$0136 appears to be significantly different from
other high-$z$ hyperluminous infrared objects that have been proven to
be type-2 objects (F~10214+4724, P~091004+4109 and F~15307+3252), it
shares several properties (continuum polarization, low ionization
broad absorption lines) with the well-known nearby ULIG Mrk~231 and
other polarized low ionization BAL quasars.

The fact that the spectrum is AGN dominated and that we do not see any
direct sign of star formation activity in the ultraviolet spectrum
does not necessarily mean that the dust that is thermally radiating in
the FIR is predominantly heated the by AGN.  Since the energy that we get
in the FIR is precisely that which is removed from the ultraviolet
spectrum, this could just mean that the starburst is more dust
enshrouded than the AGN due to a peculiar dust distribution.  Indeed,
our data allow for star formation rates as high as 2000~M$_{\sun }\,
$yr$^{-1}$.  This level of star formation is compatible with the large
amount of molecular gas infered from CO observations and with the huge
FIR luminosity of this object.

Deep spectropolarimetric observations are powerful tools to
disentangle AGN and starburst activity.  They may help to find hidden
AGN activity in putative ``pure starbursts'' galaxies. They can also
help to constrain star formation rates and starburst properties
``cleaned'' from AGN activity, independently of submillimeter
observations.  Future detailed studies of the physical properties of
more sumbmillimeter selected objects may provide important clues to
understand the link between AGN and starburst activity.

\begin{acknowledgements}
We thank Chris Lindman and Thomas Szeifert for their assistance
during the observing run at Paranal Observatory. We also thank Robert
Fosbury, Sperello di Serego Alighieri and Jacqueline Bergeron for
interesting discussions and suggestions.
\end{acknowledgements}
\bibliographystyle{apj} \bibliography{$HOME/BIB/all}
\end{document}